\documentclass[aps,prl,twocolumn,superscriptaddress,preprintnumbers,showpacs]{revtex4}
\usepackage{graphics}
\def\eq#1\en{\begin{equation} #1 \end{equation}}

\def\eqa#1\ena{\begin{eqnarray} #1 \end{eqnarray}}

\usepackage{graphicx,color}

\begin{document}

\title{Absorption of solar radiation by solar neutrinos}
\author{G. Duplan\v ci\' c}
\affiliation{Theory Division, CERN, CH-1211 Geneva 23, Switzerland}
\affiliation{Theoretical Physics Division, Rudjer Bo\v skovi\' c Institute, 
Zagreb, Croatia}
\author{P. Minkowski}
\affiliation{Theory Division, CERN, CH-1211 Geneva 23, Switzerland}
\affiliation{Institute for Theoretical Physics, University of Bern,
CH-3012 Bern, Switzerland}
\author{J. Trampeti\'{c}}
\affiliation{Theory Division, CERN, CH-1211 Geneva 23, Switzerland}
\affiliation{Theoretical Physics Division, Rudjer Bo\v skovi\' c Institute, 
Zagreb, Croatia}
\affiliation{Universit\"{a}t M\"{u}nchen, Sektion Physik, Theresienstr. 37, 80333 M\"{u}nchen, Germany}
\date{\today}

\begin{abstract}
We calculate the absorption probability of  
photons radiated from the surface of the Sun by a left-handed neutrino with definite mass
and a typical momentum for which we choose $|p_1|=0.2$ MeV,
producing a heavier right-handed antineutrino.
Considering two transitions the $\nu_1 \to \nu_2$ and $\nu_2 \to \nu_3$ 
we obtain two oscillation lengths $L_{12}=4960.8$ m, $L_{23}=198.4$ m, two absorption probabilities  
$P_{12}^{\rm abs.}\,=\,2.5\times 10^{-67}$, $P_{23}^{\rm abs.}\,=\,1.2\times 10^{-58}$  
and the two absorption ranges
$R_{12}^{\rm abs.}\,=\,4.47\times 10^4 \;R_{\odot}=208.0$ au, 
$R_{23}^{\rm abs.}\,=\,0.89\times 10^4 \;R_{\odot}=41.4$ au,
using a neutrino mass differences of $\sqrt{|\Delta m^2_{12}|}=10$ meV,  
$\sqrt{|\Delta m^2_{23}|}=50$ meV and associated transition dipole moments.
We collect all necessary 
theoretical ingredients, i.e. neutrino mass and mixing scheme, 
induced electromagnetic transition dipole moments,
quadratic charged lepton mass asymmetries and their interdependence.
\end{abstract}


\maketitle

The purpose of this paper is to determine the absorption probability of  
photons radiated from the surface of the Sun by a left-handed neutrino with definite mass,
which produces a heavier right-handed antineutrino.
To reach this goal, we start with the geometrical and thermal properties of surface radiation of the Sun 
to determine the photon flux. The elementary absorption cross section involves transition electric and 
magnetic dipole moments of neutrinos taken to be Majorana particles, in a minimal extension 
of the Standard Model to account for neutrino mass and mixing \cite{bp,hm}. 
The electromagnetic transition dipole moments are to be evaluated on the appropriate mass shells
for the neutrino-antineutrino transition. As such, they are by definition gauge invariant and
must be formed generally combining the vertex and the propagator Green functions \cite{IL}. 

Interestingly, in a recent paper \cite{sowieso}, the authors consider the fate of an electron neutrino, 
undergoing the analogous transition in the interior of the sun, induced by the interior magnetic field.
They couple this transition with normal oscillations to produce $\overline{\nu}_e$ from $\nu_e$, to be
subsequently observed on earth.

To be definite, we consider the absorption cross section
of a photon inducing an electromagnetic dipole transition from the initial
definite mass eigenstate $\nu_{L,j}(p_1)$ with mass $m_{{\nu_L},j} = m_{j}$,
with left-handed helicity,  thus denoted neutrino,
to a right-handed mass eigenstate  ${\overline\nu}_{R,k}(p_2)$ 
with mass $m_{{\nu_R},k} = m_{k}$, denoted antineutrino.
This yields the absorption probability per neutrino when integrated from the solar surface to infinity.
The kinematics of the dipole transition is then generaly the following:
\begin{eqnarray}
\gamma (q)  +  \nu_{ L,j}(p_1) \longrightarrow  \overline{\nu}_{R,k}(p_2),
\label{1}
\end{eqnarray}
and $\Delta m_{jk} =  m_{k} - m_{j}  > 0$, with $j<k\ (=1,2,3)$.
We assume hierarchical masses: $m_1 \ll m_2 \ll m_3$ and to simplify this kinematics, 
we set $m_{1} \ = \ 0$ throughout the paper.

\begin{figure}
\begin{center} 
 \resizebox{0.4\textwidth}{!}{%
  \includegraphics{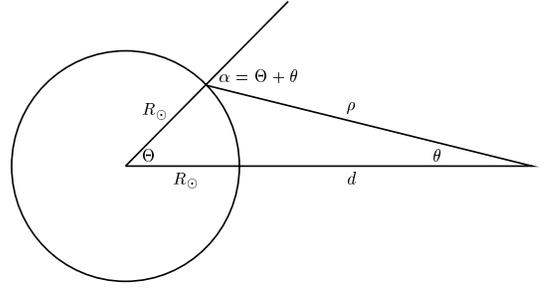}}
 \caption{Geometric properties underlying the absorption
            reaction in the reaction plane.
            The emerging left-handed neutrino is at a distance
            $R_{\odot} + d$ from the Sun's center, illuminated
            by photons emitted at an angle $\alpha$ relative
            to the normal of the emitting solar surface element.
            Thus, the triangle with the angles $\theta$, $\Theta$ and $\alpha$
            is formed.}
 \label{fig1}
 \end{center}
\end{figure}

The absorbing neutrino ($\nu_{L,j}$) is taken to emerge from
the center of the Sun, defining the radial direction as shown in
Fig. \ref{fig1}. From the geometry indicated in Fig. \ref{fig1}
we obtain the integral flux in the solar rest frame at the time when the neutrino
$\nu_{L,j}(p_1)$ is at a distance $d$ from the surface of the Sun:
\begin{eqnarray}
\phi&=&2\pi R_{\odot}^2\int_{1/a}^1 d\cos\Theta 
\label{2} \\
&\times&\int_0^{\infty}\frac{q^2dq}{4\pi^3}\frac{n_K (T)}{\rho^2}
\cos\alpha\,\cos\theta (1-\cos\theta),
\nonumber 
\end{eqnarray}
where $a = 1 + d/R_{\odot}$, $\alpha = \Theta +\theta$, with 
$R_{\odot}\cos\Theta +\rho \cos\theta =R_{\odot}+d$ and 
$ \rho \sin\theta=R_{\odot}\sin\Theta$.
The above flux represents the Planckian surface radiation of the Sun with
the temperature $T$ and the occupation number $n_K(T)$ given as $n_K(T)=(e^{K/T}-1)^{-1}$.

The differential flux multiplied by the absorption cross section $\sigma$ yields the absorption rate
per unit time $\Gamma_d(\nu_{L,j} \;\gamma \rightarrow \bar\nu_{R,k})$:
\begin{eqnarray}
\Gamma_d(\nu_{L,j} \;\gamma \rightarrow \bar\nu_{R,k}) = -\frac{d}{dt}\,\log P = \int \sigma d\phi,
\label{3}
\end{eqnarray}
where $P$ represents the survival probability of the initial neutrino $\nu_{L,j}(p_1)$.
For $c=1$, we have $t=d$, starting the clock ($t=0$) when the neutrino
$\nu_{L,j}(p_1)$ crosses the surface. The integral survival probability $P_{\infty}$ is thus
given by
\begin{eqnarray}
P_{\infty} = e^{-\int^{\infty}_0 \Gamma_{d=t} dt} = 1- \,P_{\rm abs.}\, ,
\label{4}
\end{eqnarray}
where $P_{\rm abs.}$ is the total probability for the absorption of photons.

Next, we define the absorption probability within a distance $d$ as
\begin{eqnarray}
P_{\rm abs.}(d) = 1- \,e^{-\int^{d}_0 \Gamma_{d'=t} dt} \simeq \int^{d}_0 \;\Gamma_{d'=t} dt,
\label{5}
\end{eqnarray}
because it turns out, as expected, that $P_{\rm abs.}(d) \,\leq \,P_{\rm abs.}\ll 1$.

The absorption cross section $\sigma$ is given by
\begin{eqnarray}
\sigma_{jk} = 2\pi \,|\Delta m^2_{jk}|\, \frac{|\mu_{jk}|^2}{2p_1q}
\delta\left(\cos\theta-\left(1-\frac{\Delta m^2_{jk}}{2p_1q}\right)\right),
\label{6}
\end{eqnarray}
where $|\Delta m^2_{jk}|=|m_{\nu_k}-m_{\nu_j}|^2$ and $|\mu_{jk}|$ is the 
magnetic transition dipole moment of neutrino.

Because of the purely left-handed neutrino emitted by the Sun,
there appears no spin average over neutrino helicities, contrary to photon
polarizations.

The neutrino dipole moments are determined from the effective Standard Model 
photon--neutrino vertex  
$\Gamma_{\mu}^{\rm eff}(\gamma\nu\bar\nu)$ \cite{mst,SM,dsm,MNS,FY,FKO,hnr,VV,VVO}. 
The transition $\nu_j \longrightarrow \nu_k\,\gamma$ is an electroweak process 
induced at leading 1--loop order. This order involves
the so-called ``neutrino-penguin'' diagrams through the exchange of $\ell=e,\mu,\tau$
leptons and weak bosons, and is given by \cite{IL,SM,FY}
\begin{eqnarray}
&&\epsilon^{\mu}(q)\Gamma_{\mu}^{\rm eff}(\gamma\nu\bar\nu)=\epsilon^{\mu}(q) 
\left[G_1(q^2) {\bar \nu_k}(\gamma_{\mu}q^2-q_{\mu}{\not \!q})\nu_{jL} 
\right.\nonumber \\
&&\left.+iG_2(q^2)\left( m_{\nu_k}{\bar \nu_k}\sigma_{\mu\nu}q^{\nu}\nu_{jL} + 
m_{\nu_j}{\bar \nu_k}\sigma_{\mu\nu}q^{\nu}\nu_{jR}\right)\right].
\label{7}
\end{eqnarray}
The above vertex is invariant under the electromagnetic gauge transformations.
The first term in (\ref{7}) vanishes identically for a real photon due to the electromagnetic gauge condition. 
The expression (\ref{7}) yields the electric and magnetic dipole moments \cite{SM,FY}
\begin{eqnarray}
d^{\rm el}_{kj}&=&\frac{1}{2}\left( m_{\nu_j}-m_{\nu_k}\right) \;G_2(0)_{kj},
\label{8} \\
\mu_{kj}&=&\frac{1}{2}\left( m_{\nu_j}+m_{\nu_k}\right)\;G_2(0)_{kj},
\label{9}
\end{eqnarray}
\begin{eqnarray}
G_2(0)_{kj}=\frac{2e}{M^{*2}}\hspace{-2mm}\sum_{\ell=e,\mu,\tau}{\rm U}^{\dagger}_{ki}{\rm U}^{}_{ij}\,
{\rm F}\left(x_{\ell_i}\right),\;\;x_{\ell_i}=\frac{m^2_{\ell_i}}{m^2_W},
\label{10}
\end{eqnarray}
where $i,j,k=1,2,3$ denotes neutrino species, and  
\begin{eqnarray}
{\rm F}\left(x_{\ell_i}\right)\simeq -\frac{3}{2}+\frac{3}{4}x_{\ell_i},
\label{11}
\end{eqnarray}
was obtained after the loop integration and for $x_{\ell_i}\ll 1$.
Here $M^*=4\pi\,v=3.1$ TeV, and $v=(\sqrt 2\, G_F)^{-1/2}=246$ GeV 
represents the vacuum expectation value of the scalar Higgs field. 

Note that for the off-diagonal transition moments, the first term in (\ref{11}) 
vanishes in the summation over $\ell$ due to the orthogonality condition of the mixing matrix U.

For the Majorana neutrinos considered here this matrix is  
approximative unitary and necessarily of the form
\begin{eqnarray}
\sum_{i=1}^3 {\rm U}^{\dagger}_{ki}{\rm U}^{}_{ij} = {\delta}_{kj} - \varepsilon_{kj},
\label{12}
\end{eqnarray}
where $\varepsilon$ is a hermitian nonnegative matrix (i.e. with all eigenvalues nonnegative) and 
\begin{eqnarray}
|\varepsilon|= \sqrt{{\rm Tr}\;\varepsilon^2} &=& {\cal O} \; (m_{\nu_{\rm light}}/m_{\nu_{\rm heavy}}),
\nonumber \\
&\sim& 10^{-22} \;\, {\rm to}\;\, 10^{-21}. 
\label{13}
\end{eqnarray}
The case $|\varepsilon|=0$ is excluded by the very existence of oscillation effects.
As a consequence, there is no exact GIM cancellation in lepton-flavour space, 
unlike for the cases off-diagonal transition moments and the quark-flavours.

In the Majorana case calculation of the ``neutrino--antineutrino--penguin'' diagrams,
using charged lepton and an antilepton propagators in the loops, produces
transition matrix elements which are complex antisymmetric quantities in lepton-flavour space.
Finally, 
\begin{eqnarray}
d^{\rm el}_{jk}&=&\frac{1}{2}\left( m_{\nu_j}-m_{\nu_k}\right) \;2\,{\rm Re}\;G_2(0)_{kj},
\label{14} \\
\mu_{jk}&=&\frac{1}{2}\left( m_{\nu_j}+m_{\nu_k}\right)\,2\,{\rm i}\,{\rm Im}\;G_2(0)_{kj}.
\label{15}
\end{eqnarray}
Dipole moments describing the transition from Majorana neutrino mass eigenstate-flavour 
$\nu_j$ to $\nu_k$ have the following form \cite{hm,IL,mst,MNS}:
\begin{eqnarray}
d^{\rm el}_{jk}\hspace{-1.5mm}&=&\hspace{-1.5mm}\frac{3\,e}{2M^{*2}} 
\left( m_{\nu_j}-m_{\nu_k}\right)\hspace{-2mm}
\sum_{\ell=e,\mu,\tau}\hspace{-1mm}\frac{m^2_{\ell_i}}{m^2_W}
\hspace{.5mm}{\rm Re}{\rm U}^{\dagger}_{ki}{\rm U}^{}_{ij},
\label{16}\\
{\mu}_{jk}\hspace{-1.5mm}&=&\hspace{-1.5mm}\frac{3\,e\,{\rm i}}{2M^{*2}}
\left( m_{\nu_j}+m_{\nu_k}\right)\hspace{-2mm} 
\sum_{\ell=e,\mu,\tau}\hspace{-1mm}\frac{m^2_{\ell_i}}{m^2_W} 
\hspace{.5mm}{\rm Im}{\rm U}^{\dagger}_{ki}{\rm U}^{}_{ij} .
\label{17}
\end{eqnarray}
The quadratic charged lepton mass asymmetry
is generated by the electric and magnetic dipole form factors only.
The quantities ${\rm U}_{ij}$ incorporate the neutrino-flavour mixing matrix 
governing the decomposition of a coherently
produced left-handed neutrino $\widetilde{\nu}_{L,\ell}$ 
associated with charged-lepton-flavour $\ell = e, \mu, \tau$ into
the mass eigenstates $\nu_{L,i}$:
\begin{eqnarray}
|\widetilde{\nu}_{L,\ell};\,\vec p\,\rangle =
\sum_i {\rm U}_{\ell i} |\nu_{L,i};\,\vec p ,m_i\,\rangle .
\label{18}
\end{eqnarray}
We emphasise that the sensitivity of the dipole moments (\ref{16}), (\ref{17}) is much larger because of 
the $\tau$--loop, than in the oscillations where only the mixing angles and mass differences(-square),
$|\Delta m^2_{13}| \simeq |\Delta m^2_{23}|$, enters.

We proceed to numerical evaluations
of the transition dipole moments which in general receive very small contributions 
because of the smallness of the neutrino mass, $|m_{\nu}|\sim 10^{-2}$ eV.
For the dipole moments the dominant contributions are coming from the $1 \to 2$ and $2 \to 3$ transitions. 
Since the mixing matrix element $|{\rm U}_{1\tau}|$ is small \cite{Huber}, and today
still unknown, in the evaluation of the $1 \to 2$ transition we assume the dominance of the $\mu$--loops:
\begin{eqnarray}
{|d^{\rm el}_{12}| \choose |\mu_{12}|} {\simeq} \;\frac{3e}{2M^{*2}}\;
\frac{m^2_{\mu}}{m^2_W}\sqrt{|\Delta m_{12}^2|}
{|{\rm Re}{\rm U}^{\dagger}_{1\mu}{\rm U}^{}_{\mu 2}|\choose |{\rm Im}{\rm U}^{\dagger}_{1\mu}{\rm U}^{}_{\mu 2}|}.
\label{19}
\end{eqnarray}
The dominant contributions to the electric and magnetic transition dipole moments of neutrinos 
for the $2 \to 3$ transition are, due to the $\tau$-loops, proportional 
to Re and Im part of ${\rm U}^{\dagger}_{2\tau}{\rm U}^{}_{3\tau}$ and given by
\begin{eqnarray}
{|d^{\rm el}_{23}| \choose |\mu_{23}|} {\simeq} \;\frac{3e}{2M^{*2}}\;
\frac{m^2_{\tau}}{m^2_W}\sqrt{|\Delta m_{23}^2|}
{|{\rm Re}{\rm U}^{\dagger}_{2\tau}{\rm U}^{}_{\tau 3}|\choose |{\rm Im}{\rm U}^{\dagger}_{2\tau}{\rm U}^{}_{\tau 3}|}.
\label{20}
\end{eqnarray}
Setting for the matrix elements $|{\rm Im}{\rm U}^{\dagger}_{1\mu}{\rm U}^{}_{\mu 2}| \simeq 0.32$ and
$|{\rm Im}{\rm U}^{\dagger}_{2\tau}{\rm U}^{}_{\tau 3}| \simeq 0.5$ \cite{MNS,FKO}, 
and specifically for hierarchical masses 
$\sqrt{|\Delta m^2_{12}|}\simeq 10\,\rm meV$ \cite{SNO} and 
$\sqrt{|\Delta m^2_{32}|}\simeq 50\,\rm meV$ \cite{Kamio} 
and transforming the moments to Bohr magneton units, we have found the following standard transition magnetic moments 
of neutrinos
\begin{eqnarray}
|\mu_{12}|_{\rm st} &\simeq& 3.12 \times 10^{-34} \;\rm eV^{-1} = 1.05 \times 10^{-27}\, \mu_{\rm B},
\label{21} \\
|\mu_{23}|_{\rm st}  &\simeq& 6.14 \times 10^{-31} \;\rm eV^{-1} = 2.07\times 10^{-24}\, \mu_B.
\label{22}
\end{eqnarray}
From above equations we see that the transition $1 \to 3$ is sensitive to 
value of $|{\rm U}_{1\tau}|$ down to the ${\cal O}(10^{-3})$ \cite{Huber}.
Note finally that direct experimental evidence for neutrino flavour transformation from 
neutral-current interactions (\ref{7}) is given in Ref. \cite{SNO1}.

The absorption rate per unit time in the solar rest system at a given distance $d$ 
from the Sun's surface is 
\begin{eqnarray}
&&\Gamma_{jk}^d(\nu_{L,j} \;\gamma \rightarrow \bar\nu_{R,k}) =\frac{1}{2\pi^2} \int q^2\;dq\,
d\cos\theta \,n_K(T)  
\nonumber\\
&&\times\;\sigma_{jk}(q,\theta)\;\cos\theta\;(1-\cos\theta)
\frac{\sin\Theta \cos(\theta + \Theta)}{\sin\theta(a\cos\Theta -1)}.
\label{23}
\end{eqnarray}
In Eq. (\ref{23}) the factor ${\sin\Theta \cos(\theta + \Theta)}/
{\sin\theta(a\cos\Theta -1)}$ is equal to 1, as a consequence of the geometry described in  Fig. 1.

From Eqs. (\ref{6}), (\ref{21}) and (\ref{22}), we obtain the standardized form of the absorption cross section
\begin{eqnarray}
\sigma_{\rm st} = X_{jk}\;\sigma_{jk}^{\rm red.}=\frac{X_{jk}}{2p_1q}
\delta\left(\cos\theta-\left(1-\frac{\Delta m^2_{jk}}{2p_1q}\right)\right),
\label{24}
\end{eqnarray}
where $\sigma_{jk}^{\rm red.}$ represents the reduced cross section and $X_{jk}$ is dimensionless quantity
\begin{eqnarray}
X_{jk}=2\pi \;|{\Delta m_{jk}^2}|\; |\mu_{jk}|^2_{st},
\label{25}
\end{eqnarray}
which receive the following values for the respective magnetic moments (\ref{21}), (\ref{22}) and mass differences:
\begin{eqnarray}
X_{12}&=&6.10\times 10^{-71},
\label{26}\\
X_{23}&=&5.91\times 10^{-63}.
\label{27}
\end{eqnarray}
The reduced cross section $\sigma_{jk}^{\rm red.}$ leads to 
the following reduced dimensionless absorption rate $\Gamma_{jk}^{\rm red.}$:
\begin{eqnarray}
\Gamma_{jk}^{\rm red.}(a) &=&\frac{1}{2\pi^2} R_{\odot}\int q^2\;dq\,
d\cos\theta \,n_K(T)
\nonumber\\
&\times& \sigma_{jk}^{red.}(q,\theta) \,(1-\cos\theta)\cos\theta .
\label{28}
\end{eqnarray}
The integral 
\begin{eqnarray}
f_{jk}(A)=\int_1^A da\;\Gamma_{jk}^{\rm red.}(a),\;\;\; A=1+\frac{d}{R_{\odot}}
\label{29}
\end{eqnarray}
determines the absorption probability defined by Eq. (\ref{5}) 
within a distance $d$ from the solar surface
\begin{eqnarray}
P_{jk}^{\rm abs.}(d)=X_{jk}\,f_{jk}(A),\;\; P_{jk}^{\rm abs.}=X_{jk}\,f_{jk}(\infty).
\label{30}
\end{eqnarray}

The radius of the Sun given in different units is
\begin{eqnarray}
R_{\odot} =6.961\times 10^8 \;\rm m= 2.322\;\rm s = 3.528\times 10^{15}\;\rm eV^{-1}.
\label{31}
\end{eqnarray}
The astronomical unit (Sun--Earth distance) corresponds to
$\rm au\equiv {\oplus}-{\odot}=499.005 \;\rm s$, which means that
$A_{\oplus}={\rm au}/{R_{\odot}} = 214.91$. 

\begin{figure}
 \begin{center}
 \resizebox{0.4\textwidth}{!}{%
  \includegraphics{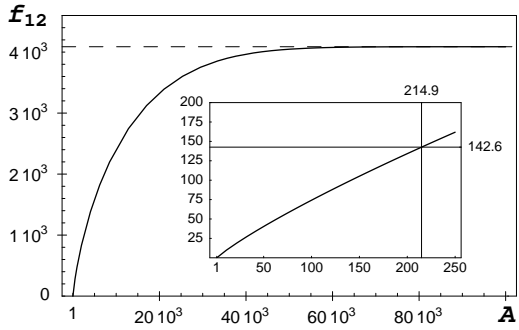}}
 \caption{The absorption probability $f_{12}(A)$ within a distance $d$ from the surface of the Sun as a 
 function of $A=1+d/R_{\odot}$.}
 \label{fig2}
 \end{center}
\end{figure}

\begin{figure}
 \begin{center}
 \resizebox{0.4\textwidth}{!}{%
  \includegraphics{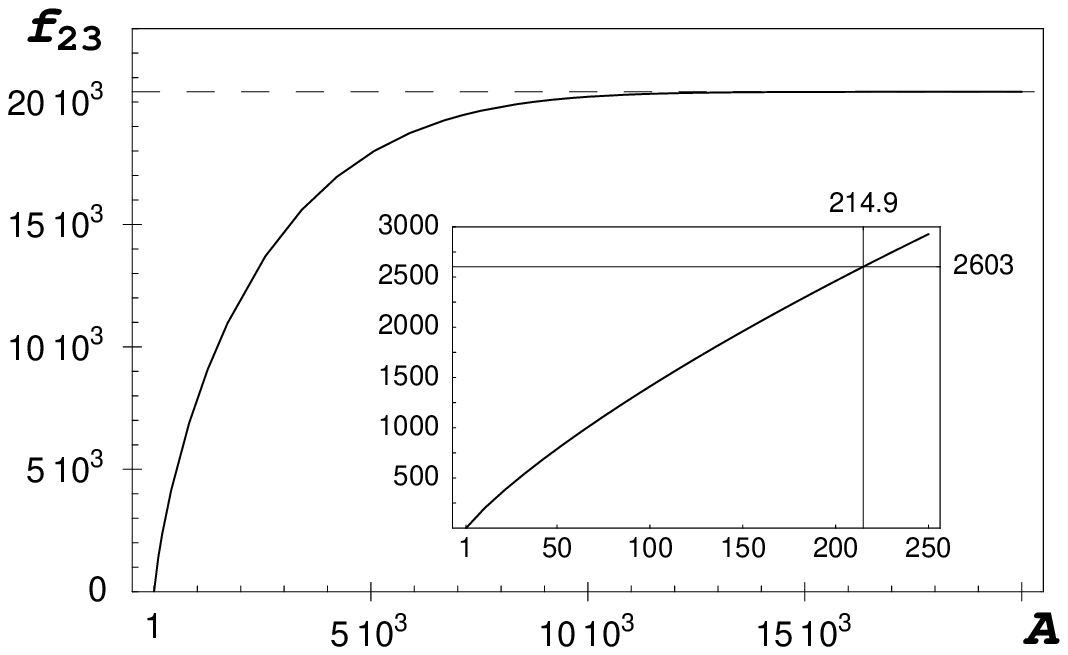}}
 \caption{The absorption probability $f_{23}(A)$ within a distance $d$ from the surface of the Sun as a 
 function of $A=1+d/R_{\odot}$.}
 \label{fig3}
 \end{center}
\end{figure}

We choose $p_1=0.2$ MeV for the momentum of the solar neutrino emerging from 
the solar interior through the surface \cite{as},
yielding the oscillation lengths $L_{12}$ and $L_{23}$ of neutrinos $\nu_1 \longleftrightarrow \nu_2$
and $\nu_2 \longleftrightarrow \nu_3$, respectively:
\begin{eqnarray}
L_{12}\hspace{-2mm}&=\hspace{-1mm}&\frac{4\pi p_1}{|\Delta m^2_{12}|}= 25.133\times \, 10^{9} \,
\rm eV^{-1}= 4960.8 \;\rm m,
\label{32}\\
L_{23}\hspace{-2mm}&=\hspace{-1mm}&\frac{4\pi p_1}{|\Delta m^2_{23}|}= 1.005\times \, 10^{9} \,
\rm eV^{-1}= 198.4 \;\rm m.
\label{33}
\end{eqnarray}
The solar activity is characterized by the surface temperature $T_{\odot}$, which we take as
0.5 eV = 5802.5 K. We note that a significant contribution to the absorption probability comes from 
wave--lengths of solar radiation comparable with the oscillation lengths $L_{12}$ and $L_{23}$. Those
long wave--lengths constitute the Rayleigh--Jeans tail of the Planckian spectrum.
For the standards defined above, we present the functions $f_{12}(A)$ and $f_{23}(A)$ in  
Figs. \ref{fig2} and \ref{fig3}, respectively. The enlarged scale figures are
displayed separately up to $\simeq$ 1 au.

The analytic structure of the functions $f_{12}(A)$ and $f_{23}(A)$ gives rise 
to a characteristic range of solar activity in the absorption process,
independently of the small absorption probability:
\begin{eqnarray}
R_{12}^{\rm abs.}\hspace{-2mm}&=&\hspace{-2mm}R_{\odot}\,\sqrt{\frac{2p_1 T_{\odot}}{|\Delta m^2_{12}|}} =
4.47\times 10^4 R_{\odot}=208.0\;\rm au,
\label{34}\\
R_{23}^{\rm abs.}\hspace{-2mm}&=&\hspace{-2mm}R_{\odot}\,\sqrt{\frac{2p_1 T_{\odot}}{|\Delta m^2_{23}|}} =
0.89\times 10^4 R_{\odot}=41.4\;\rm au.
\label{35}
\end{eqnarray}
The absorption range $R_{23}^{\rm abs.}$, for example 
corresponds to a distances of 2 astronomical units beyond the 
semi--major axis of the orbit of Pluto. Ranges $R_{12}^{\rm abs.}$, $R_{23}^{\rm abs.}$ are clearly
visible near $A=5\times 10^4$ in Fig. \ref{fig2}, in the approach of the function
$f_{12}(A)$ to its asymptotic value $f_{12}(\infty)=4087.9$,
and near $A=10^{4}$ in Fig. \ref{fig3}, in the approach of the function
$f_{23}(A)$ to its asymptotic value $f_{23}(\infty)=20418.8$.
Thus the total absorption probabilities per neutrino become
\begin{eqnarray}
P_{12}^{\rm abs.}\,&=&\,X_{12}\,f_{12}(\infty)\,\simeq\ 2.5 \times 10^{-67},
\label{36}\\
P_{23}^{\rm abs.}\,&=&\,X_{23}\,f_{23}(\infty)\,\simeq\,1.2 \times 10^{-58}.
\label{37}
\end{eqnarray}
For the bound on the neutrino magnetic moment,  
$\mu_{\nu_e}\,\stackrel{<}{\sim}\,3\times 10^{-12} \mu_B$, derived from SN1987A \cite{AOT},
we would obtain $P_{abs.}\sim  10^{-35}$. Taking $10^{40}$ similar neutrinos, emitted from 
the pp cycle of the sun, the rate of so produced antineutrinos from the sun would be $10^5$ per year. 
Our number (\ref{36}) is meant as a lower bound.

In this work we collected all necessary 
theoretical ingredients, i.e. neutrino mass and mixing scheme, 
induced electromagnetic transition dipole moments,
quadratic mass asymmetries entering the process 
of photon absorption by a definitive neutrino flavor 
producing a massive antineutrino.
Obviously the same electromagnetic dipole transitions occur under many more circumstances,
e.g. in the core of supernova explosions, in reactions with the cosmic microwave background radiation, 
in strong magnetic fields like in neutron stars, 
in coherent Maser light directed at a nuclear reactor. These processes may well lead to larger effects. 

As follows from the geometric aspects discussed in Figs. 1, 2 and 3 and displayed in Eqs. (\ref{30})--(\ref{37}),
the absorption probability per neutrino scales from our solar example to another similar
situation, modula minor logarithmic correction factors, as 
\begin{eqnarray}
\frac{R\;\,\langle T\rangle }{p^2},
\label{38}
\end{eqnarray}
where $R_{\odot}$ is replaced by an irradiating photon surface of radius $R$, $\langle T\rangle$
denotes an average temperature of surface of light emission, replacing $T_{\odot}=0.5$ eV,
and $p$ denotes the momentum of the emitted neutrino. 
From the above scaling law it follows, that no dramatic changes in absorption probability
(per neutrino or antineutrino) occur in a supernova explosion. There the momenta are
changed to ${\cal O}$(10-20 MeV) from the neutrinosphere, $R$ is adapted to the size of source,
before the neutrino is beyond the range of its radiation, i.e. cannot be many orders of
magnitude larger than the solar radius, and $\langle T\rangle$ is (much) below the MeV scale,
when becoming effective in the process of $\nu \longrightarrow \bar\nu$ dipole transition.
The only difference is in the number of neutrinos emitted, offset by the 
distance(-square) attenuation, towards an observer on earth. For this reason we refrain from
giving numerical estimates for supernova explosions here.

\vspace{1cm}

The work of P.M. is supported by the Swiss National Science Foundation.
The work of G.D. and J.T. is supported by the Ministry of Science and Technology 
of Croatia under Contract No. 0098002.

\end{document}